\documentclass[a4paper,10pt,twoside]{cpc-hepmad}

\usepackage{multicol}
\usepackage{graphicx}
\usepackage{booktabs}
\usepackage{amssymb,bm,mathrsfs,bbm,amscd}
\usepackage[tbtags]{amsmath}
\usepackage{lastpage}

\begin{document}

\fancyhead[co]{\footnotesize F. Jugeau: On the
non-perturbative corrections for
$\Lambda_b\rightarrow\Lambda_c\ell\bar{\nu}_{\ell}$ in Heavy
Quark Effective Theory}

\title{On the non-perturbative corrections for $\Lambda_b\rightarrow\Lambda_c\ell\bar{\nu}_{\ell}$ in Heavy Quark Effective Theory\thanks{Invited talk at the 5$^{th}$ High-Energy Physics International Conference HEP-MAD 11, Antananarivo, Madagascar, 25-31$^{th}$ August 2011.}}

\author{%
Fr\'ed\'eric Jugeau%
} \maketitle

\address{%
TPCSF, Institute of High Energy Physics, Chinese Academy of
Sciences, Beijing 100049, China\\
jugeau@ihep.ac.cn
}

\begin{abstract}
We study consequences of the non-forward amplitude for the baryon
decay $\Lambda_b\rightarrow\Lambda_c\ell\bar{\nu}_{\ell}$
which will be measured in detail at $LHCb$. We obtain a sum rule
for the subleading elastic Isgur-Wise (IW) function $A(w)$ that
originates from the kinetic part of the $O(1/m_Q)$ effective
Lagrangian perturbation. In the sum rule appear only the
intermediate states $J_j^P=\frac{1}{2}_0^+$, the same that
contribute to the $O(1/m^2_Q)$ correction to the axial-vector form
factor $G_1(w)$ involved in the differential decay rate at zero
recoil $w=1$. This allows us to obtain a lower bound on the
correction $-\delta^{(G1)}_{1/m^2_Q}$ in terms of $A(w)$ and the
shape of the leading elastic IW function $\xi_{\Lambda}(w)$.
Another theoretical implication is that $A'(1)$ must vanish in the
limit where the slope $\rho_{\Lambda}$ of $\xi_{\Lambda}(w)$
saturates its lower bound. A strong correlation between the
leading IW function $\xi_{\Lambda}(w)$ and the subleading one
$A(w)$ is thus established in the case of the baryons.
\end{abstract}



\begin{multicols}{2}

\section{Introduction}

We aim at investigating within the Heavy Quark Effective Theory
formalism (HQET) the $O(1/m_Q^2)$ subleading corrections to the
baryonic semileptonic decay
$\Lambda_b\rightarrow\Lambda_c\ell\bar{\nu}_{\ell}$\cite{jugeau}. An
important ingredient is the consideration of the non-forward
amplitude
$\Lambda_b(v_i)\rightarrow\Lambda_c(v')\rightarrow\Lambda_b(v_f)$
allowing for general velocities $v_i$, $v_f$ and $v'$. At leading
order, the HQET sum rule method has been applied to the case of
the baryonic leading elastic IW function
$\xi_{\Lambda}(w)=1-\rho_{\Lambda}^2(w-1)+\frac{\sigma_{\Lambda}^2}{2}(w-1)^2+\ldots$,
giving the following lower bounds for its slope\cite{IWY1991} and
its curvature\cite{LOR2009}:
\begin{align}
\rho^2_{\Lambda}&=\sum_{n\geq0}[\tau_1^{(n)}(1)]^2\geq0\;,\label{slope bound}\\
\sigma_{\Lambda}^2&=\frac{3}{5}\left(\rho_{\Lambda}^2+(\rho_{\Lambda}^2)^2+\sum_{n\neq0}[\xi_{\Lambda}^{(n)\prime}(1)]^2\right)\geq\frac{3}{5}\Big(\rho_{\Lambda}^2+(\rho_{\Lambda}^2)^2\Big)\label{curvature bound}
\end{align}
where $\tau_1^{(n)}(1)$ denote the IW functions of transition
$j^P=0^+\rightarrow1^-$ at zero recoil ($j^P$ is the spin-parity
of the so-called \emph{brown muck}, namely, all the light degrees
of freedom within the heavy-light baryon, and $n$ is a radial
quantum number).

This leading as well as the new subleading results presented in
this letter consist of a set of constraints that the differential
decay rate of $\Lambda_b\rightarrow\Lambda_c\ell\bar{\nu}_{\ell}$ -
which will be accurately measured at $LHCb$ - must fulfill. Data
has already been obtained for this mode at the Tevatron giving a
large branching ratio of about 5\% and a large fraction of the
inclusive semileptonic decay
$BR(\Lambda_b\rightarrow\Lambda_c\ell\bar{\nu}_{\ell}+\textrm{anything})\simeq10\%$.

The proceeding is organized as follows. We first
introduce the leading and subleading IW functions
$\xi_{\Lambda}(w)$ and $A(w)$ relevant for the $O(1/m_Q^2)$
non-perturbative correction $\delta^{(G_1)}_{1/m_Q^2}$ to the
axial-vector form factor $G_1(1)$ at zero recoil which enters into
the expression of the differential rate of the semileptonic decay
$\Lambda_b\rightarrow\Lambda_c\ell\bar{\nu}_{\ell}$. In
Sections 3 and 4, we derive sum rules respectively for the
subleading elastic IW function $A(w)$ and for the correction
$\delta^{(G_1)}_{1/m_Q}$. Section 5 presents a bound for the
latter and, in Section 6, we discuss the strong correlation
between leading and subleading baryonic quantities in HQET.

\section{Leading and subleading baryonic IW functions}

Of particular interest for us will be the form factor $G_1(w)$
which enters into the parametrization of the baryonic matrix
element of the axial-vector current\cite{FNbaryon1990}:
\begin{multline}\label{definition of G}
\langle\Lambda_c(v',s')\mid\bar{c}\gamma^{\mu}\gamma_5
b\mid\Lambda_b(v,s)\rangle=\\
\bar{u}_{\Lambda_c}(v',s')\Big[G_1(w)\gamma^{\mu}+G_2(w)v^{\mu}+G_3(w)v^{\prime\mu}\Big]\gamma_5
u_{\Lambda_b}(v,s)
\end{multline}
where $w\equiv v\cdot v'$ is the recoil and $u_{\Lambda_Q}(v,s)$
is the spinor of the heavy baryon physical state such that
${\not{v}}u_{\Lambda_Q}(v,s)=u_{\Lambda_Q}(v,s)$ and
$\bar{u}_{\Lambda_Q}(v,s)u_{\Lambda_Q}(v,r)=2m_{\Lambda_Q}\delta_{rs}$
($p_{\Lambda_Q}^{\mu}=m_{\Lambda_Q}v^{\mu}$ with $m_{\Lambda_Q}$
the physical mass of the heavy-light baryon $\Lambda_Q$).

Indeed, in the neighborhood of the zero recoil $w=1$ kinematic
point, the differential rate of the transition
$\Lambda_b\rightarrow\Lambda_c\ell\bar{\nu}_{\ell}$ depends
only on $G_1(1)$: \begin{equation}\label{differential rate}
\frac{1}{\sqrt{w^2-1}}\frac{d\Gamma}{dw}\underset{w\simeq1}{\simeq}\frac{G_F^2\mid
V_{cb}\mid^2}{4\pi^3}m^3_{\Lambda_c}(m_{\Lambda_b}-m_{\Lambda_c})^2\mid
G_1(1)\mid^2\,.
\end{equation}
In the heavy quark expansion, the form factor $G_1(w)$ is
expressed, at the order $O(1/m_Q)$ ($Q=b$ or $c$ quark), in terms
of two IW functions $\xi_{\Lambda}(w)$ and $A(w)$:
\begin{equation}\label{HQETexpansion}
G_1(w)=\xi_{\Lambda}(w)+\left(\frac{1}{2m_b}+\frac{1}{2m_c}\right)\Big[\frac{w-1}{w+1}\bar{\Lambda}\xi_{\Lambda}(w)+A(w)\Big]\,.
\end{equation}
The leading elastic IW function $\xi_{\Lambda}(w)$ is defined by
the matrix element of the lowest-order heavy-heavy current
$J=\bar{h}_{v'}^{(Q')}\Gamma h^{(Q)}_v$ in HQET ($\Gamma$ is
any combination of gamma matrices):
\begin{multline}
\langle\Lambda_c(v',s')\mid\bar{h}_{v'}^{(c)}\Gamma
h^{(b)}_v\mid\Lambda_b(v,s)\rangle=\\
\xi_{\Lambda}(w)\,\bar{\mathcal{U}}_{\Lambda_c}(v',s')\Gamma\,\mathcal{U}_{\Lambda_b}(v,s)
\end{multline}
where
$\mathcal{U}_{\Lambda_Q}(v,s)=\left(1+O(1/m_Q^2)\right)^{-1/2}u_{\Lambda_Q}(v,s)$
is the spinor of the effective heavy baryon state in HQET
normalized such that
$\bar{\mathcal{U}}_{\Lambda_Q}(v,s)\mathcal{U}_{\Lambda_Q}(v,r)=2M_{\Lambda_Q}\delta_{sr}$.
The effective mass of the HQET state
$M_{\Lambda_Q}=m_Q+\bar{\Lambda}$ is given in terms of the
energy $\bar{\Lambda}$ of the \emph{brown muck}.
$h_v^{(Q)}(x)=e^{im_Q v\cdot
x}\left(\frac{1+\not{v}}{2}\right)Q(x)$ is the effective heavy
quark field which appears when building the effective Lagrangian
as a power series expansion in $1/m_Q$:
\begin{equation}
\mathcal{L}^{(Q)}_{eff}=\mathcal{L}^{(Q)}_{HQET,v}+\mathcal{L}^{(Q)}_{kin,v}+\mathcal{L}^{(Q)}_{mag,v}
\end{equation}
with\footnote{The perpendicular component of the covariant
derivative $D$ is defined by $D_{\mu\perp}=D_{\mu}-(v\cdot D)
v_{\mu}$. Also, $[D_{\mu},D_{\nu}]=ig_s G_{\mu\nu}$ and
$\sigma_{\mu\nu}=\frac{i}{2}[\gamma_{\mu},\gamma_{\nu}]$.}
\begin{equation}
\left\{\begin{array}{l}
\mathcal{L}^{(Q)}_{HQET}=\bar{h}^{(Q)}_v(iv\cdot
D)h^{(Q)}_v\;,\\
\mathcal{L}^{(Q)}_{kin,v}=\frac{1}{2m_Q}\mathcal{O}^{(Q)}_{kin,v}\;\;;\;\;\mathcal{O}^{(Q)}_{kin,v}=\bar{h}^{(Q)}_v(iD_{\perp})^2h^{(Q)}_v\;,\\
\mathcal{L}^{(Q)}_{mag}=\frac{1}{2m_Q}\mathcal{O}^{(Q)}_{mag,v}\;\;;\;\;\mathcal{O}^{(Q)}_{mag,v}=-\frac{g_s}{2}\bar{h}^{(Q)}_v\sigma_{\mu\nu}G^{\mu\nu}h^{(Q)}_v\;.
\end{array}
\right. \end{equation} 

As for $A(w)$, it stems from the kinetic
part of the $O(1/m_Q)$ effective Lagrangian:

\begin{multline}
\langle\Lambda_c(v',s')\mid i\int d^4x
T\{J(0),\mathcal{O}_{kin}^{(Q)}(x)\}\mid\Lambda_b(v,s)\rangle=\\
A(w)\,\bar{\mathcal{U}}_{\Lambda_c}\,\Gamma\,
\mathcal{U}_{\Lambda_b}\;.\label{definition of A}
\end{multline}

It is worth pointing out that, contrary to the case of the meson
ground-state doublet $j^P=\frac{1}{2}^-$ where the transition
matrix element of $B\rightarrow
D^{(\ast)}\ell\bar{\nu}_{\ell}$ gets two \emph{Current}-type
and three \emph{Lagrangian}-type $O(1/m_Q)$ corrections (the
latter coming both from $\mathcal{L}_{kin}$ and
$\mathcal{L}_{mag}$)\cite{FNmeson1990,J1,J2}, in the case of the
baryon ground-state singlet $j^P=0^+$ considered here, $G_1$ gets
only one \emph{Current}-type corrections (in terms of
$\bar{\Lambda}\xi_{\Lambda}$) and only one
\emph{Lagrangian}-type correction (in terms of $A(w)$) as the
magnetic part $\mathcal{O}^{(Q)}_{mag,v}$ of the $O(1/m_Q)$
Lagrangian perturbation doesn't contribute to
$A(w)$\cite{GGW1990}.

Furthermore, due to the vector current conservation, we have the
condition $A(1)=0$ at zero recoil\cite{FNbaryon1990}.
Consequently, according to \eqref{HQETexpansion}, the form factor
$G_1(1)$ in \eqref{differential rate} has, at zero recoil, only
$O(1/m_Q^2)$ corrections which we will denote as follows:
\begin{equation}\label{corrections definition}
 G_1(1)=1+\delta^{(G_1)}_{1/m_Q^2}\,.
\end{equation}
The main goal of the present letter is to study these corrections\cite{jugeau}.

\section{HQET sum rule for the subleading IW function $A(w)$}

Since only the kinetic perturbation $\mathcal{O}^{(Q)}_{kin}$ of the
effective Lagrangian contributes to the definition
\eqref{definition of A} of $A(w)$, only the radially excited
intermediate states $\Lambda_c^{(n)}$ with the same spin-parity
$J_j^P=\frac{1}{2}_0^+$ as the ground-state $\Lambda_c$ are
relevant and give us the following sum rule:
\begin{equation}\label{sum rule A}
A(w)=\sum_{n\neq0}\frac{\xi_{\Lambda}^{(n)}(w)}{\Delta
E^{(n)}}\frac{\langle\Lambda_c^{(n)}(v,s)\mid\mathcal{O}^{(c)}_{kin,v}(0)\mid\Lambda_c(v,s)\rangle}{\sqrt{4m_{\Lambda_c^{(n)}}m_{_{\scriptstyle\Lambda_c}}}\sqrt{v^0_{\Lambda_c^{(n)}}v^0_{_{\scriptstyle\Lambda_c}}}}
\end{equation}
where $\Delta E^{(n)} \equiv m_{\Lambda_c^{(n)}} - m_{_{\scriptstyle\Lambda_c}}$. Especially, we recover $A(1)=0$ since
$\xi_{\Lambda}^{(n)}(1)=\delta_{n,0}$.

\section{HQET sum rule for the correction $\delta^{(G_1)}_{1/m_Q^2}$ to the form factor $G_1(1)$}

Analogously to the sum rule formulated in the mesonic case for the
axial-vector current (Eqns (114) and (5.6) of \cite{BSUV1995} and
\cite{LLSW1998} respectively), we have for the form factor $G_1$
at the order $O(1/m_Q^2)$:
\end{multicols}
%\ruleup
\begin{equation}
\mid
G_1(1)\mid^2+{1 \over 2}\sum_{s,s'} \sum_{n\not= 0}{\mid\langle \Lambda_c^{(n)}(0^+,1^+)(v,s')\mid \vec{A}\mid\Lambda_b(0^+)(v,s) \rangle\mid^2 \over 4m_{\Lambda_c^{(n)}}m_{_{\scriptstyle\Lambda_b}}}=1+\Big[\left(\frac{1}{2m_c}-\frac{1}{2m_b}\right)^2+\frac{8}{3}\frac{1}{2m_c}\frac{1}{2m_b}\Big]\lambda
\end{equation}
{\flushleft{where the hard radiative corrections have been
neglected and where
$-\lambda=\frac{1}{2M_{\Lambda_b}}\langle\Lambda_b(v)\mid\bar{h}_v^{(b)}(iD_{\perp})^2h_v^{(b)}\mid\Lambda_b(v)\rangle$
is the mean kinetic energy value. The $O(1/m_Q^2)$ correction to
$G_1$ takes then the following expression:}}
\begin{align}\label{sum rule correction}
-\delta^{(G_1)}_{1/m_Q^2}=&-\frac{1}{2}\Big[\left(\frac{1}{2m_c}-\frac{1}{2m_b}\right)^2+\frac{8}{3}\frac{1}{2m_c}\frac{1}{2m_b}\Big]\lambda\nonumber\\
&+ {1 \over 4} \sum_{s,s'}  \sum_{n\not=0}{\mid\langle \Lambda_c^{(n)}(0^+)(v,s') \mid\vec{A}\mid \Lambda_b(v,s) \rangle\mid^2 \over 4m_{\Lambda_c^{(n)}}m_{_{\scriptstyle\Lambda_b}}} 
+ {1 \over 4}  \sum_{s,s'} \sum_{n\not= 0}{\mid\langle \Lambda_c^{(n)}(1^+)(v,s') \mid\vec{A}\mid \Lambda_b(v) \rangle\mid^2 \over 4m_{\Lambda_c^{(n)}}m_{_{\scriptstyle\Lambda_b}}}\,. 
\end{align}
\ruledown \vspace{0.5cm}
\begin{multicols}{2}
{\flushleft{The matrix elements implicitly contain the double
insertion (on the initial $b$- and final $c$-legs) of the kinetic
and magnetic parts of the $O(1/m_Q)$ effective Lagrangian to the
HQET axial-vector current
$A^{\mu}=\bar{h}_v^{(c)}\gamma^{\mu}\gamma_5h_v^{(b)}$ for
which only the spatial component $\overrightarrow{A}$ survives in
the heavy quark limit $m_Q\rightarrow\infty$. In \eqref{sum rule
correction}, the final states
$\Lambda_c^{(n)}(J^P_j=\frac{1}{2}_0^+)$ and
$\Lambda_c^{(n)}(J_j^P=\frac{1}{2}_1^+,\frac{3}{2}_1^+)$ are
attained respectively by $\mathcal{L}_{kin}$ and
$\mathcal{L}_{mag}$. For the sake of argument, we have:}}

\end{multicols}
\ruleup
\begin{align}
\langle \Lambda_c^{(n)}(v,s') \mid\bar{h}_v^{(c)}\vec{A}h_v^{(b)}(0)\mid \Lambda_b(v,s)\rangle&={1 \over 2m_c} \langle\Lambda^{(n)}_c(v,s') \mid i\int d^4xT \{\mathcal{O}^{(c)}_{kin,v}(x),\bar{h}_v^{(c)}\vec{A}h_v^{(b)}(0)\}\mid \Lambda_b(v,s)\rangle\\
&+\ {1 \over 2m_b} \langle\Lambda^{(n)}_c(v,s') \mid i \int d^4xT \{\mathcal{O}^{(b)}_{kin,v}(x),\bar{h}_v^{(c)}\vec{A}h_v^{(b)}(0)\}\mid \Lambda_b(v,s) \rangle
\end{align} 
which, by inserting the intermediate states and using the
flavor-spin heavy quark symmetry, finally gives:
\begin{equation}
\langle \Lambda_c^{(n)}(v,s') \mid\bar{h}_v^{(c)}\vec{A}h_v^{(b)}(0)\mid \Lambda_b(v,s) \rangle\underset{O(1/m_Q)}{=}\frac{-1}{\Delta E^{(n)}}\left(\frac{1}{2m_c}-\frac{1}{2m_b}\right)\frac{\langle\Lambda_c^{(n)}(v,s)\mid\mathcal{O}^{(c)}_{kin,v}(0)\mid \Lambda_c(v,s)\rangle}{\sqrt{4m_{\Lambda_c^{(n)}}m_{_{\scriptstyle\Lambda_c}}}\sqrt{v^0_{\Lambda_c^{(n)}}v^0_{_{\scriptstyle{\Lambda_c}}}}}\mathcal{U}_{\Lambda_c}^\dagger(v,s') \vec{\Sigma}\mathcal{U}_{\Lambda_b}(v,s)
\end{equation}
{\flushleft{where $\overrightarrow{\Sigma}=\textrm{diag}(\overrightarrow{\sigma},\overrightarrow{\sigma})$
is the double Pauli matrix.}}

A similar evaluation of the second matrix element in \eqref{sum
rule correction} between the ground-state singlet
$\Lambda_b(\frac{1}{2}_0^+)$ and the excited states
$\Lambda_c^{(n)}(\frac{1}{2}_1^+,\frac{3}{2}_1^+)$ gives a
positive semi-definite contribution and, in particular, a term
proportional to $1/(4m_cm_b)$ reminiscent of the one found in\cite{FNbaryon1990}:
\begin{equation}
G_1(1)=1+\frac{1}{2}\Big[\left(\frac{1}{2m_c}-\frac{1}{2m_b}\right)^2+\frac{8}{3}\frac{1}{2m_c}\frac{1}{2m_b}\Big]\lambda
+\frac{1}{2}\left(\frac{1}{2m_c}-\frac{1}{2m_b}\right)^2[-D_1(1)+3D_2(1)]+\frac{1}{2m_c}\frac{1}{2m_b}4D_2(1)
\end{equation}
where the functions $D_i$ ($i=1,2$) correspond to the double
insertion of the operators $\mathcal{O}_{kin,v}^{(Q)}$ and
$\mathcal{O}_{mag,v}^{(Q)}$ on the initial and final heavy quark
legs ($Q=b,c$).

\section{Bound on the corrections $\delta^{(G_1)}_{1/m_Q^2}$}

The preceding analysis allows us to write a bound for the
$O(1/m_Q^2)$ corrections to $G_1(1)$. Eqn. \eqref{sum rule
correction} implies indeed that
\begin{equation}\label{correction bound}
- \delta_{1/{m_Q^2}}^{(G_1)} \geq - {1 \over 2} \left [ \left ({1 \over 2m_c}  - {1 \over 2m_b} \right )^2 + {8 \over 3} {1 \over 2m_c} {1 \over 2m_b} \right ] \lambda
+ {1 \over 2} \left({1 \over 2m_c}  - {1 \over 2m_b} \right)^2 \sum_{n\not=0} \left[{1 \over \Delta E^{(n)}} {\langle\Lambda_c^{(n)}(v,s)\mid\mathcal{O}^{(c)}_{kin,v}(0)\mid\Lambda_c(v,s)\rangle \over \sqrt{4m_{\Lambda_c^{(n)}}m_{_{\scriptstyle\Lambda_c}}}\sqrt{v^0_{\Lambda_c^{(n)}}v^0_{_{\scriptstyle\Lambda_c}}}}\right] ^2\;.
\end{equation}
The crucial point here is that the intermediate states entering
into \eqref{correction bound} are the same that the ones
contributing to the sum rule \eqref{sum rule A} for $A(w)$. Using
the Cauchy-Schwarz inequality
$\mid\sum_{n}A_nB_n\mid^2\leq\left(\sum_n\mid
A_n\mid^2\right)\left(\sum_n\mid B_n\mid^2\right)$, the latter
gives: 
\begin{equation}
\left [ A(w) \right ]^2 \leq\sum_{n\not=0} \left [ \xi^{(n)}_\Lambda (w)\right ]^2 \sum_{n\not= 0}
\left [{1 \over \Delta E^{(n)}} {\langle\Lambda_c^{(n)}(v,s)\mid\mathcal{O}_{kin, v}^{(c)}(0)\mid\Lambda_c(v,s)\rangle \over \sqrt{4m_{\Lambda_c^{(n)}}m_{_{\scriptstyle\Lambda_c}}}\sqrt{v^0_{\Lambda_c^{(n)}}v^0_{_{\scriptstyle\Lambda_c}}}}\right ]^2
\end{equation}
such that
\begin{equation}
-\delta_{1/{m_Q^2}}^{(G_1)}\geq - {1 \over 2} \left [ \left ({1 \over 2m_c}  - {1 \over 2m_b} \right )^2 + {8 \over 3} {1 \over 2m_c} {1 \over 2m_b} \right ] \lambda +\ {1 \over 2} \left ( {1 \over 2m_c} - {1 \over 2m_b} \right )^2 \frac{[A(w)]^2}{\sum_{n\not=0} \left[ \xi^{(n)}_\Lambda (w)\right]^2}\;.
\end{equation}
Since this inequality is valid for any value of $w$, we can
consider its limit $w\rightarrow1$, taking into account that
$A(1)=0$ and $\xi_{\Lambda}^{(n)}(1)=\delta_{n,0}$:
\begin{equation}
- \delta_{1/{m_Q^2}}^{(G_1)} \geq  - {1 \over 2} \left[ \left({1 \over 2m_c}  - {1 \over 2m_b} \right)^2 + {8 \over 3} {1 \over 2m_c} {1 \over 2m_b} \right] \lambda +\ {1 \over 2} \left( {1 \over 2m_c} - {1 \over 2m_b}\right)^2 {[A'(1)]^2 \over \sum_{n\not=0} \left[ \xi^{(n)}_\Lambda{'}(1)\right]^2}
\end{equation}
and from \eqref{curvature bound},
$\displaystyle{\sum_{n\neq0}[\xi_{\Lambda}^{(n)\prime}(1)]^2=\frac{5}{3}\sigma_{\Lambda}^2-\rho_{\Lambda}^2-(\rho_{\Lambda}^2)^2}$,
one finally gets:
\begin{equation}\label{main result}
- \delta_{1/{m_Q^2}}^{(G_1)}\geq - {1 \over 2} \left[ \left({1 \over 2m_c}  - {1 \over 2m_b} \right)^2 + {8 \over 3} {1 \over 2m_c} {1 \over 2m_b} \right] \lambda +\ {3 \over 10} \left( {1 \over 2m_c} - {1 \over 2m_b}\right)^2 {[A'(1)]^2 \over \sigma_\Lambda^2 - {3 \over 5}[\rho_\Lambda^2+(\rho_\Lambda^2)^2]}
\end{equation}
\ruledown \vspace{0.5cm}
\begin{multicols}{2}
{{\flushleft which is the main result of the proceeding.}} From
Eqns \eqref{differential rate} and \eqref{corrections definition},
we see that the $O(1/m_Q^2)$ correction at zero recoil
$-\delta^{(G_1)}_{1/m_Q^2}$ is pivotal in the extrapolation of the
semileptonic differential decay rate
$\Lambda_b\rightarrow\Lambda_c\ell\bar{\nu}_{\ell}$ near zero
recoil. In particular, this is needed to check that the value of
$\mid V_{cb}\mid$ that would fit exclusive baryon semileptonic
data is indeed consistent with what we presently know on this
parameter from the meson exclusive and inclusive determinations.
It is in this respect that the bound \eqref{main result} is
important.

\section{Correlation between $A'(1)$ and the shape of the leading elastic IW function $\xi_{\Lambda}(w)$}

Since the inequality \eqref{main result} holds for any values of
$\rho_{\Lambda}^2$ and $\sigma_{\Lambda}^2$ satisfying the
constraints \eqref{slope bound} and \eqref{curvature bound}, it
should hold for their lowest values. However, if the curvature
attains its lowest value
$\sigma_{\Lambda}^2\rightarrow\frac{3}{5}\left(\rho_{\Lambda}^2+(\rho_{\Lambda}^2)^2\right)$,
the second term on the \emph{r.h.s.} of \eqref{main result} would
diverge. Because this behavior is unphysical, we predict instead a
strong correlation between $A'(1)$ and the shape of the leading
elastic IW function $\xi_{\Lambda}(w)$. Eqn. \eqref{main result}
implies the correlation:
\begin{equation}\label{first correlation}
\textrm{if}\;\;\;\;\sigma_{\Lambda}^2\Rightarrow\frac{3}{5}\left(\rho_{\Lambda}^2+(\rho_{\Lambda}^2)^2\right)\;\;\;\;\textrm{then}\;\;\;\;A'(1)\Rightarrow0\,.
\end{equation}
As a matter of fact, a group theory-based method
(equivalent to the HQET sum rule approach) has been developed to
study the baryonic IW functions\cite{LOR2010}. It turns out that if the
slope of $\xi_{\Lambda}(w)$ attains its lowest possible value
$\rho_{\Lambda}^2=0$, then all the higher-order derivatives
$\xi_{\Lambda}^{(n)}(1)$ ($n\geq2$) vanish at zero recoil.
Especially, that implies $\sigma_{\Lambda}^2=0$. In view of
\eqref{first correlation}, we have then:
\begin{equation}\label{second correlation}
\textrm{if}\;\;\;\;\rho_{\Lambda}^2\Rightarrow0\;\;\;\;\textrm{then}\;\;\;\;A'(1)\Rightarrow0\,.
\end{equation}
The non-trivial results \eqref{first correlation} and
\eqref{second correlation} relate the behavior of the leading
elastic IW function $\xi_{\Lambda}(w)$ to the subleading effective
form factor $A(w)$.\\
\\
\acknowledgments{I am grateful to the organizers of MAD-HEP 11,
and especially to its Chairman S. Narison, who gave me the
opportunity to present my research activities on the Effective Theories of QCD during this
conference.}

\end{multicols}

\vspace{-2mm}
\centerline{\rule{80mm}{0.1pt}}
\vspace{2mm}

\begin{multicols}{2}

\end{multicols}

\clearpage

\end{document}